\documentstyle[aas2pp4]{article}
\lefthead{Schlichenmaier, Jahn, Schmidt}
\righthead{Penumbral fine structure and the Evershed effect}

\begin{document}

\title{A dynamical model for the penumbral fine structure \\
and the Evershed effect in sunspots}

\author{R. Schlichenmaier\altaffilmark{1}}
\affil{Max-Planck-Institut f\"ur extraterrestrische Physik,
85748 Garching, Germany}

\author{K. Jahn}
\affil{Warsaw University Observatory, Al. Ujazdowskie 4, 00--478
Warsaw, Poland; crj@astrouw.edu.pl}

\and

\author{H.U. Schmidt}
\affil{Max-Planck-Institut f\"ur Astrophysik, 85748 Garching, Germany}

\altaffiltext{1}{present address: Kiepenheuer-Institut f\"ur
Sonnenphysik, Sch\"oneckstr. 6, 79104 Freiburg, Germany; schliche@kis.uni-freiburg.de}

\begin{abstract}

 Relying on the assumption that the interchange convection of magnetic flux 
 tubes is the physical cause for the existence of sunspot penumbrae, we
 propose a model in which the dynamical evolution of a thin magnetic flux
 tube reproduces the Evershed effect and the penumbral fine structure such
 as bright and dark filaments and penumbral grains.

 According to our model, penumbral grains are the manifestation of the 
 footpoints of magnetic flux tubes, along which hot subphotospheric plasma 
 flows upwards with a few km/s.  Above the photosphere the hot plasma
 inside the tube is cooled by radiative losses as it flows horizontally
 outwards. As long as the flowing plasma is hotter than the surroundings,
 it constitutes a bright radial filament.  The flow confined to a thin
 elevated channel reaches the temperature equilibrium with the surrounding
 atmosphere and becomes optically thin near the outer edge of the penumbra.
 Here,  the tube has a height of approximately 100 km above the continuum
 and the flow velocity reaches up to 14 km/s. Such a flow channel can
 reproduce the observed signatures of  the Evershed effect.

\end{abstract}


\section{Introduction}

 A sunspot penumbra reveals quite a number of phenomena that are not yet
 fully understood. In high resolution images, high intensity contrasts
 that evolve dynamically are observed: Radially elongated bright and
 dark filaments, having a width $<$ 0.35$''$ (Muller
 1973b\markcite{mul73b}), and penumbral grains migrating towards the
 umbra. Muller describes penumbral grains as having a {\it comet-like}
 structure: a bright {\it coma} and a somewhat thinner and dimmer tail
 pointing radially outwards.  There is observational evidence that
 bright filaments consist of a few penumbral grains, that are radially
 aligned (Muller 1973b\markcite{mul73b}; Soltau 1982\markcite{sol82};
 see also Muller 1992)\markcite{mul92}.

 Analyzing the Stokes-I profiles of photospheric absorption lines within
 the penumbra, Evershed (1909\markcite{eve09}) measured a shift of the
 line core, indicating a radial and horizontal outflow of
 mass. Additional characteristic features of the Evershed effect are
 that penumbral Stokes-I profiles are asymmetric, and that the line
 shift is most intense in dark filaments (e.g. Beckers \& Schr\"oter
 1969\markcite{bec69}; Degenhardt 1993\markcite{deg93}; Wiehr
 1995\markcite{wie95}). Stationary siphon-flow models (Meyer \& Schmidt
 1968\markcite{mey68}) seem promising in order to explain the Evershed
 effect (Thomas 1988\markcite{tho88}; Degenhardt 1991\markcite{deg91};
 Thomas \& Montesinos 1993\markcite{tho93}). However, these models
 depend on the existence of strong magnetic flux concentrations near the
 outer edge of the penumbra, in order to create a gas pressure gradient
 between two footpoints of an arched magnetic flux tube that accelerates
 the plasma. Here, we want to present a first step towards a dynamical
 model, in which the gas pressure gradient that accelerates the plasma
 builds up locally within the penumbra.

 It has been pointed out by Schmidt (1987\markcite{HUS87}, see also 
 Schmidt 1991\markcite{HUS91}) that the penumbra containing a significant
 fraction of the sunspot's magnetic flux has to be deep in the sense that
 it extends many scale heights beneath the photosphere. In order to explain
 the surplus brightness of the penumbra as compared to the umbra, Jahn \&
 Schmidt (1994, hereafter JS) proposed the concept of interchange
 convection, in which heat that leaks through the magnetopause is
 distributed horizontally within the penumbra by an interchange of magnetic
 flux tubes. Parameterizing the effects of interchange convection, JS
 constructed a tripartite model that consists of three distinct 
 hydrostatic stratifications for the umbra, the penumbra, and the quiet
 sun, respectively.  Two current sheets, i.e., the peripatopause separating
 the umbra from the penumbra and the magnetopause separating the penumbra
 from the quiet sun, create a current-free magnetostatic field that is
 consistent with the observed geometry of the magnetic field, and which
 satisfies total pressure equilibrium horizontally across the current
 sheets.

 Studying the consequences of interchange convection, we perform numerical
 simulations which describe the dynamical evolution of a thin magnetic flux
 tube embedded in a tripartite sunspot model of JS  (sec. 2; for some
 preliminary results see also Jahn et al. 1996, and Schlichenmaier et al.
 1997a). Our numerical experiments are compared with observed penumbral
 phenomena in section 3. We argue that our dynamical model can consistently
 explain the appearance of bright and dark filaments, penumbral grains, and
 the Evershed effect, and conclude (sec. 4) that interchange convection
 offers a promising concept towards an understanding of the penumbra.

 \section{The evolution of a thin magnetic flux tube within the penumbra}

 We assume that the penumbral magnetic flux fragments into magnetic flux
 tubes. These flux tubes take part in the process of interchange
 convection, and evolve dynamically as physical entities within the
 penumbra. Thus, the penumbra consists of an ensemble of magnetic flux
 tubes. As a natural first step towards an understanding of such a
 penumbra, we study the dynamics of one single flux tube that is embedded
 in a tripartite sunspot model of JS as a pregiven background.

 The dynamical evolution of the magnetic flux tube is described by taking
 advantage of the thin flux tube approximation (Defouw
 1976\markcite{def76}; Moreno--Insertis 1986\markcite{mor86}).  Thus, we
 neglect magnetic diffusivity and assume that any physical variable does
 not change across the tube, but can only vary along the tube. We are left
 with a one-dimensional tube evolving in a  two-dimensional background. As
 the independent variable we use the  integrated mass, $a$, along the tube,
 and use convective time derivatives to track individual mass elements.  It
 is assumed that the tube instantaneously reaches the total pressure
 equilibrium with the background:  
 \begin{equation} p(a,t) + {B^2(a,t)\over 8\pi} 
                = p_{\rm b}(a,t)  + {B_{\rm b}^2(a,t) \over 8\pi}
                  \;\;.\label{pressure_equilibrium} 
 \end{equation} 
 Here, $p$ and $B$ denote the gas pressure and the magnetic field strength, 
 respectively, and the index b marks local background variables. The
 equation of state takes into account the partial ionization of hydrogen
 and helium. The radiative heat exchange of the tube with the background is
 described using the relaxation time approach of Spiegel
 (1957\markcite{spi57}). Details of the mathematical and numerical
 description are given in Schlichenmaier et al. (1997b\markcite{sch97b}; see
 also Schlichenmaier 1997\markcite{sch97}).

 \placefigure{fig1}

 Figure 1a shows the initial configuration of our numerical experiment. The
 tube, having a magnetic flux $\phi=2\cdot 10^{16}$ Mx, lies along the
 magnetopause, so that part of the tube is in contact with the underlying
 hotter quiet sun.  As the lower boundary condition, the lowest mass
 element of the tube is held fixed at a depth of $z=-15$ Mm and a radial
 distance from sunspot center of $x=5$ Mm, i.e. at the local position of
 the magnetopause.  At the upper end of the tube, initially at $x=25$ Mm
 and $z=400$ km, we choose free boundary conditions: The time derivative of
 the tangent vector with the modulus length per mass increment is
 extrapolated and the total force at the last interior grid point is
 transfered to the outermost grid point. Figure 1
 \notetoeditor{Figure 1 should be one-column-wide} 
 shows the peripatopause, the magnetopause, and the shape of the tube
 (with its diameter scaled by a factor 12 for better visibility). The
 photosphere of the quiet sun ($\tau=2/3$ level) corresponds to $z=0$
 km. The horizontal line at $z=-150$ km in the penumbra represents the
 Wilson-depressed photosphere of the penumbra. Figure 1a also shows the
 Wilson depression of the umbra at $z=-470$ km. In Figure 1b and 1c
 subsequent stages of the evolution of the tube are shown. The shading
 represents the flow speed along the tube. The background shading
 corresponds to vanishing velocity and any darker shading to a velocity
 that points upwards (increasing $z$) and outwards (increasing $x$).

 The part of the tube that is in contact with the underlying hotter quiet
 sun is heated by radiation, expands, gets less dense than the
 surroundings, and rises due to buoyancy. Below the photosphere, the
 ascending part of the tube is accelerated, because radiative heat exchange
 is negligible (i.e., the tube rises adiabatically) and the stratification
 is superadiabatic (i.e., convectively unstable). Above the photosphere the
 tube ceases to rise, since the tube's density increases due to radiative
 losses and due to the stabilizing background stratification. The rate at
 which the tube emerges is dominated by the magnitude of superadiabaticity
 in the background stratification, which has its maximum a few 100 km below
 the photosphere. Thus, the tube starts to rise just below the photosphere,
 and subsequently the upward motion extends to deeper layers. The footpoint
 of the tube, i.e. the intersection of the tube with the photosphere,
 migrates towards the umbra as the subphotospheric part of the tube rises
 (see Fig. 1b).

 The forces acting perpendicular to the tube are buoyancy, ${\bf\hat
 n} \cdot {\bf g} (\rho-\rho_{\rm b})$, magnetic tension, $ \kappa
 B^2/4\pi$, and the gradient of the magnetic pressure in the background, $
 -{\bf\hat n} \cdot \nabla B_{\rm b}^2 / 8\pi$ ($\rho$ and $B$ are density
 and magnetic field  strength, respectively, $\bf g$ is the gravity at the
 solar surface, $\kappa$ measures the tube's curvature, and ${\bf\hat n}$
 denotes the unit vector perpendicular to the tube).  Since the magnetic
 pressure decreases upwards inside a sunspot, the tube is pushed upwards.
 Approximately 100 km above the penumbral photosphere the tube finds  an
 equilibrium in a form of a horizontal channel (with vanishing curvature
 force) in which the anti-buoyancy force (the tube's density is 10\% larger
 than the background density) is balanced by the gradient of  the
 background's magnetic pressure directed upwards.

 The tube stays in total pressure equilibrium with the background as it
 rises (cf. eq. \ref{pressure_equilibrium}). The magnetic background
 pressure decreases very little relative to the background gas pressure,
 which decreases exponentially with a scale height of the order of 100
 km.  This implies that below the photosphere the total pressure
 decreases according to the gas pressure scale height, since $\beta =
 4\pi p/B^2 \gg 1$. Thus, the plasma within the tube expands as it
 rises, and because magnetic flux is conserved along the tube, the
 magnetic pressure within the tube decreases.  From equation
 (\ref{pressure_equilibrium}) it follows that the gas pressure
 inside the tube must become larger than the gas pressure of the
 background, so that a surplus gas pressure builds up within the tube as
 it rises.

 At the outer edge of the penumbra, above the photosphere, the tube does
 not rise and the gas pressure inside the tube does not change there. The
 part of the tube that has risen to the photosphere has a surplus gas
 pressure. Since the external pressure above the photosphere is almost
 constant on horizontal planes a gas pressure gradient, $\partial p /
 \partial a$, builds up along the horizontal part of the tube. In result,
 the plasma is accelerated outwards, since gravity, ${\bf\hat t}\cdot {\bf
 g}\rho$, cannot counteract the gas pressure gradient in an horizontal
 tube. That means that as the tube rises through the subphotospheric
 penumbra, a longitudinal flow develops pointing upwards below the
 photosphere and horizontally outwards above the photosphere.

 As the inclination of the subphotospheric part of the tube becomes
 steeper, the buoyancy diminishes and the footpoint stops migrating
 inwards. Now, one would expect that magnetic tension pulls the tube
 back down again, but at the turning point the longitudinal flow exerts
 a centrifugal force that can counteract the magnetic tension and
 prevent the tube from sinking down.  Figure 1c shows the tube after
 5400 s. One can see that the footpoint of the tube, which started at
 $x=13.5$ Mm, has migrated inwards to $x\approx 8.5$ Mm. For the further
 evolution of the tube, the reader is referred to Schlichenmaier et
 al. (1997b).

 \section{Comparison with observations}

 \placefigure{fig2}

 With this simulation at hand we can offer an explanation for the penumbral
 grains, bright and dark filaments, and the Evershed effect. In Figure
 2
 \notetoeditor{Figure 2 should be two-column-wide}
 a magnified snapshot of the tube, which is elevated some 100 km above the
 penumbral photosphere, is shown for $t=5400$ s. Here the gray scale
 represents the temperature of the tube and of the background. The length
 of the arrows in the tube corresponds to the longitudinal flow speed. The
 tube's diameter is magnified by a factor of 6.

 \paragraph{Penumbral grains:}

 The upflow in the subphotospheric tube brings up hot plasma, which makes
 the footpoint look bright. At the footpoint the upflow velocity amounts to
 3 km/s. A penumbral grain can be explained by such a footpoint, because
 footpoints are hotter and brighter than the surroundings and migrate
 inwards towards the umbra. Also, due to the high temperature, the optical
 thickness of the tube that corresponds to the diameter of the tube at the 
 footpoint amounts to $\tau\approx 10^{3}$, indicating that penumbral
 grains should be optically thick. 

 \paragraph{Bright filaments:}

 From the footpoint, the hot plas\-ma flows outward horizontally and cools 
 gradually due to radiative losses. At some point (in our simulation at 
 $x\approx 10.5$ Mm, see Figure 2) the flow reaches temperature 
 equilibrium with the surroundings.

 The horizontal part of the tube that is hotter than the surroundings can
 be identified as the tail of a penumbral grain. This part is dimmer than
 the footpoint because it is cooler. It also gets compressed as the plasma
 loses internal energy, and the tube becomes thinner. Our model then
 suggests that bright filaments are the tails of penumbral grains. However,
 it may also be that an observed bright filament consists of a few
 penumbral grains representing some distinct flux tubes that are  radially
 aligned.

 \paragraph{Dark filaments:}

 As long as the plasma is hotter than the surrounding atmosphere, its
 opacity is high due to the temperature dependence of the H$^-$ opacity,
 $\tilde\kappa({\rm H}^{-}) \propto T^{9}$. Hence, the tube is optically
 thick as long as it is hot. When the flow within the tube reaches
 temperature equilibrium it becomes as transparent ($\tau \approx 10^{-1}$)
 as the atmosphere at a similar height. In our simulated tube, the length of
 the bright filament would measure $\approx 2000$ km with a thickness of
 less than 50 km. The length of a bright filament depends on how fast the
 plasma is cooled. A thinner tube would have a smaller optical thickness
 and would be cooled more efficiently, forming a shorter bright filament.

 According to our model, dark filaments do not exist {\it per se}, but are 
 caused by optically thick bright filaments that partially cover the darker
 penumbral  photosphere. The spacing between two adjacent bright filaments
 that are  radially elongated appears as a dark filament. This idea is
 consistent with the statement of  Muller (1973a\markcite{mul73a},
 1973b\markcite{mul73b}), that within the penumbra, bright features {\it
 show up} against a dark background.

\paragraph{Evershed effect:}

 Once the plasma reaches temperature equilibrium with its surroundings,
 it becomes optically thin, i.e., $\tau \approx 10^{-1}$. Hence, as can
 be seen in Figure 2, the tube is transparent between $x\approx 10.5$ Mm
 and the outer edge of the penumbra. There, a line-of-sight that crosses
 the tube reaches optical depth $\tau=2/3$ near the photosphere of the
 model atmosphere at $z\approx -150$ km. Thus, the flux tube appears as
 a flow channel, being transparent, thin, and elevated with respect to
 the photosphere. Between the footpoint and the point of temperature
 equilibrium the gas pressure inside the tube decreases due to radiative
 losses. By this gas pressure gradient, the flow is accelerated from 3
 km/s at the footpoint up to a supercritical speed of 14 km/s near the
 outer edge of the penumbra.

 In order to reproduce the characteristic features of the Evershed
 effect, namely that penumbral line profiles are shifted and asymmetric,
 such that the line wing is more shifted than the line core (see, e.g.,
 Degenhardt 1993\markcite{deg93}), we propose the following
 geometry. Assume that the line core originates mainly some 200 km above
 the penumbral photosphere, whereas the main contributions for the line
 wing (20\% line depression) stems from, say 100 km above continuum. A
 flow channel at and below 100 km above continuum would then only affect
 and shift the line wing, whereas the line core would be mostly
 unaffected. For such a configuration, our model reproduces observed
 line profiles, as, e.g., the line profile of Fe I 709.0 nm, which is
 shown in Figure 1b of Degenhardt (1993\markcite{deg93}). In this
 context, we mention that Rimmele (1995\markcite{rim95}) and Stanchfield
 et al. (1997\markcite{sta97}) interpret their observations as being
 caused by thin elevated flow channels.

 Wiehr (1995\markcite{wie95}) deduces a flow velocity of $\ge 5$ km/s by
 assuming that the line asymmetry is caused by a spatially unresolved
 line satellite caused by a flow channel. If one further takes
 into account that a typical contribution function has a full width at
 half maximum of more than 100 km and that our simulated tube has a
 diameter of less than 50 km, a local flow velocity exceeding 10 km/s is
 still consistent with observation.

\section{Discussion and conclusion}

 Relying on the concept of interchange convection, we study the dynamics
 of one thin magnetic flux tube that evolves in the penumbra. The
 results of this paper show that penumbral grains, bright filaments, and
 the Evershed effect may be caused by the dynamical evolution of
 magnetic flux tubes that are embedded in the penumbra. According to our
 model, bright penumbral phenomena are the consequence of an upflow of
 hot subphotospheric plasma that is channeled by magnetic flux tubes.
 Simultaneously, one can understand the Evershed flow as being the
 extension of that upflow. After the flow is cooled due to radiation
 losses, the tube becomes transparent. The asymmetry in the line profile
 appears because the flow channel affects mostly the line wing of a
 photospheric absorption line.

 We have studied the evolution of one particular flux tube that lies
 along the magnetopause initially. Beyong the visible edge of the
 penumbra the magnetopause ascends toward the magnetic canopy in a
 convectively stable atmosphere. In our simulation, the flow follows the
 magnetopause. Sofar, we did not study the case of a downstream leg near
 the outer edge of the penumbra.

 Our model suggests that the concept of interchange convection seems
 very promising. We can propose a coherent picture of the physical
 structure of the penumbra based on the results presented
 here. Furthermore, the concept of interchange convection seems to
 explain the brightness of the penumbra as compared to the umbra. As we
 show in this paper, an upflow along magnetic flux tubes develops,
 transporting hot subphotospheric plasma up to the photosphere. However,
 for an understanding of the behavior of a complete ensemble of
 penumbral flux tubes, different initial conditions have to be studied.

\acknowledgements
RS gratefully acknowledges the support by Prof. G. Haerendel.
KJ acknowledges the support by the KBN grant 2 P03D 010 12.

\newpage

\figcaption[series.eps]{
The shape of the flux tube is shown for $t=0,\,1200,$ and $5400$
s. Umbra and penumbra are separated by the peripatopause, penumbra and
quiet sun by the magnetopause. The horizontal line within the penumbra
represents the photosphere of the penumbra. The gray scale shows the
flow speed along the tube. Note that the tube's diameter is magnified
by a factor 12 for better visibility.
\label{fig1}}

\figcaption[temperatur.eps]{
At $t=5400$ s the tube near the solar surface is shown. The horizontal
line at $z=-150$ km represents the photosphere of the penumbra. The gray
scale represents the temperature of the tube and of its surroundings. On
the left side the cooler umbra and on the right side the hotter quiet
sun are visible.  The tube's diameter is magnified by a factor 6. The
arrows represent longitudinal flow speeds. In the lower left corner an
arrow representing 5 km/s is drawn.
\label{fig2}}

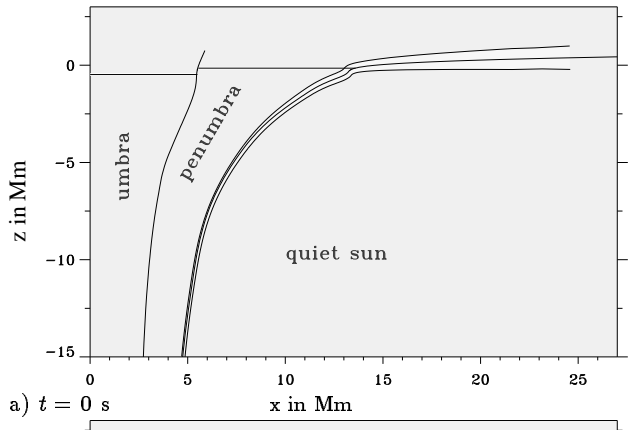
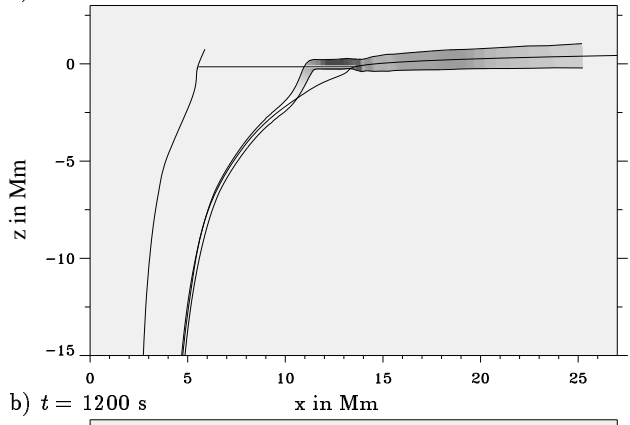
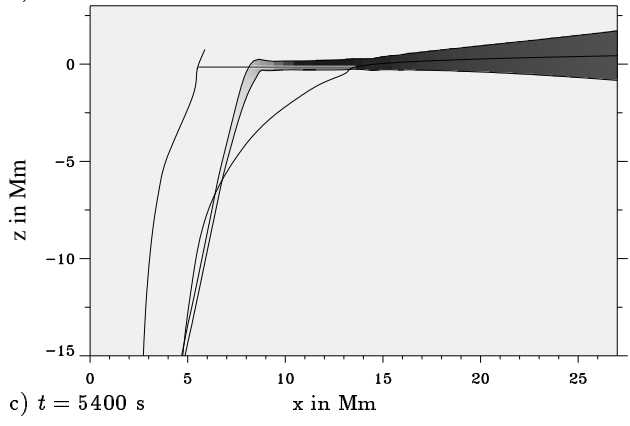
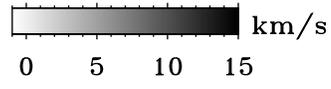
\end{document}